\documentclass[3p,times]{elsarticle}

\usepackage[utf8]{inputenc}
\usepackage[T1]{fontenc}
\usepackage{textcomp}

\usepackage{amsmath,amssymb}

\usepackage[colorlinks=true]{hyperref}

\journal{Int.~J.~Mass~Spectrom.}

\begin{document}

\begin{frontmatter}

\title{Stacked-Ring Ion Guide for Cooling and Bunching Rare Isotopes}

\author[rug-vsi]{X.~Chen\corref{xc}}
\cortext[xc]{Corresponding author}
\ead{xiangcheng.chen@rug.nl}

\author[rug-vsi]{J.~Even}
\author[greifswald]{P.~Fischer}
\author[darmstadt]{M.~Schlaich}
\author[rug-ziam]{T.~Schlathölter}
\author[greifswald]{L.~Schweikhard}
\author[rug-vsi]{A.~Soylu}

\affiliation[rug-vsi]{
    organization={Van Swinderen Institute for Particle Physics and Gravity, University of Groningen},
    addressline={Zernikelaan 25},
    city={Groningen},
    postcode={9747 AA},
    country={The Netherlands},
}
\affiliation[greifswald]{
    organization={Institute for Physics, University of Greifswald},
    addressline={Felix-Hausdorff-Straße 6},
    city={Greifswald},
    postcode={17489},
    country={Germany},
}
\affiliation[darmstadt]{
    organization={Institute for Nuclear Physics, Technical University Darmstadt},
    addressline={Schlossgartenstraße 9},
    city={Darmstadt},
    postcode={64289},
    country={Germany},
}
\affiliation[rug-ziam]{
    organization={Zernike Institute for Advanced Materials, University of Groningen},
    addressline={Nijenborgh 4},
    city={Groningen},
    postcode={9747 AG},
    country={The Netherlands},
}

\begin{abstract}
In modern rare isotope facilities, ion cooling and bunching lies at the heart of the ion transfer along a low-energy beam line that consists of several differential pumping stages.
We present a conceptual design of an ion guide as an alternative to the conventional linear Radio-Frequency Quadrupole (RFQ) for cooling and bunching rare isotopes.
The ion guide is composed of stacked ring electrodes of varying apertures, to which a confining RF potential following a rectangular waveform is applied.
The thicknesses of the rings and the gaps in between are varied accordingly to maximize the confining volume and to reduce ion losses.
Ion transport within the ion guide is facilitated by a lower-frequency wave traveling on top of the higher-frequency confining field.
The former is induced by locally adjusting the duty cycle of the rectangular waveform of the confining potential.
Design parameters are first calculated by analytical studies and then optimized by ion trajectory simulations with SIMION\textsuperscript\textregistered.
The results show that the ion guide enables high ion transmission and produces well focused ion bunches.
It will be used in the NEXT project---an experimental study of atomic masses of Neutron-rich EXotic nuclei produced in multi-nucleon Transfer reactions.
\end{abstract}


\end{frontmatter}

\section{Introduction}
Rare isotope facilities have opened up unprecedented opportunities for atomic and nuclear researches on exotic nuclei that lie on the outskirts of the nuclide chart~\cite{kluge_atomic_2010}.
Being complementary to the Isotope Separation On-Line (ISOL) method, in-flight separation is proved to be a chemically independent technique to access rare isotopes with an excess proton-to-neutron imbalance~\cite{blumenfeld_facilities_2013}.
When coupled to a buffer-gas cooling section, such as a gas-filled stopping cell~\cite{wada_genealogy_2013}, high-quality beams of rare isotopes can be prepared and transferred to ion traps for precision measurements or to post-accelerators for further experiments.

Ions extracted from the low-vacuum gas cell need to be confined transversely while being transferred to the next differential pumping stage.
The transverse confinement is usually provided by a linear Radio-Frequency Quadrupole (RFQ)~\cite{herfurth_segmented_2003} amid an inert buffer gas, typically being helium owing to its high ionization energy and light mass, at a pressure between $10^{-3}$ mbar and $10^{-1}$ mbar~\cite{douglas_collisional_1992}.
Apart from the ion cooling due to the dissipative ion-neutral collisions, ion bunching is frequently needed to efficiently inject ions into traps or post-accelerators.
RFQ-type ion guides are used as an ion cooler and buncher in many rare isotope facilities across the globe.
One straightforward version is axially segmented for the application of appropriate Direct-Current (DC) potentials to the segments to store ions in the axial direction~\cite{herfurth_linear_2001,nieminen_beam_2001,darius_linear_2004,neumayr_ion-catcher_2006,lunney_colette_2009,brunner_titans_2012,beyer_rfq_2014,mehlman_status_2015,barquest_development_2016,boussaid_technical_2017}.
An improved segmentation, which diagonally cuts every rod of an RFQ into half, can result in less wiring~\cite{barquest_rfq_2017,valverde_cooler-buncher_2020}.
Inspired by enclosing an RF sextupole with three ring electrodes for ion bunching~\cite{fujitaka_accumulation_1997}, several groups instead added wedge electrodes to efficiently provide an axial guiding force along an RFQ with reduced DC potentials~\cite{podadera_aliseda_design_2004,traykov_compact_2011,schwarz_lebit_2016,ricketts_compact_2020}.
Ion cooling and ion bunching can also be realized by two individual but coupled RFQs~\cite{traykov_production_2008,haettner_versatile_2018}, which can offer more flexibility, such as an orthogonal ion ejection~\cite{ito_novel_2013}.

Despite the worldwide success of RFQs, their phase-dependent acceptances necessitate carefully matching the incoming beam emittances~\cite{dawson_quadrupole_1976}.
Otherwise, the effective acceptance is drastically limited.
A stacked-ring ion guide~\cite{guan_stacked-ring_1996}, which possesses a rotational symmetry, intrinsically avoids this matching problem, and is thus particularly suitable for a cylindrically symmetric beam extracted from, e.g., a circular orifice of a gas cell. 
A notable example is the ion funnel~\cite{kelly_ion_2010}, which typically features a wide aperture at the entrance to capture a divergent beam and gradually shrinks its aperture towards the exit to focus the beam~\cite{neumayr_ion-catcher_2006,brunner_rf-only_2015,querci_rf-only_2018,varentsov_windowless_2020}.
Sophisticated boundary shapes can be adopted to serve for specific purposes, such as ion trapping inside~\cite{ibrahim_ion_2007}.

Furthermore, the RF potential applied to an ion cooler and buncher is not limited to the conventional harmonic type with a sinusoidal waveform.
``Digital'', i.e.\ rectangularly shaped, RF potentials have recently received much attention~\cite{ding_ion_2006,lee_simulation_2011,bandelow_stability_2013,bandelow_3-state_2013,neuwirth_using_2015,xu_design_2016,yu_characterizing_2019}, as they provide stronger transverse confinement and facilitate adaptation across a broad frequency band~\cite{brunner_titans_2012}.

\begin{figure}[htbp]
    \centering
    \includegraphics[width=.9\textwidth]{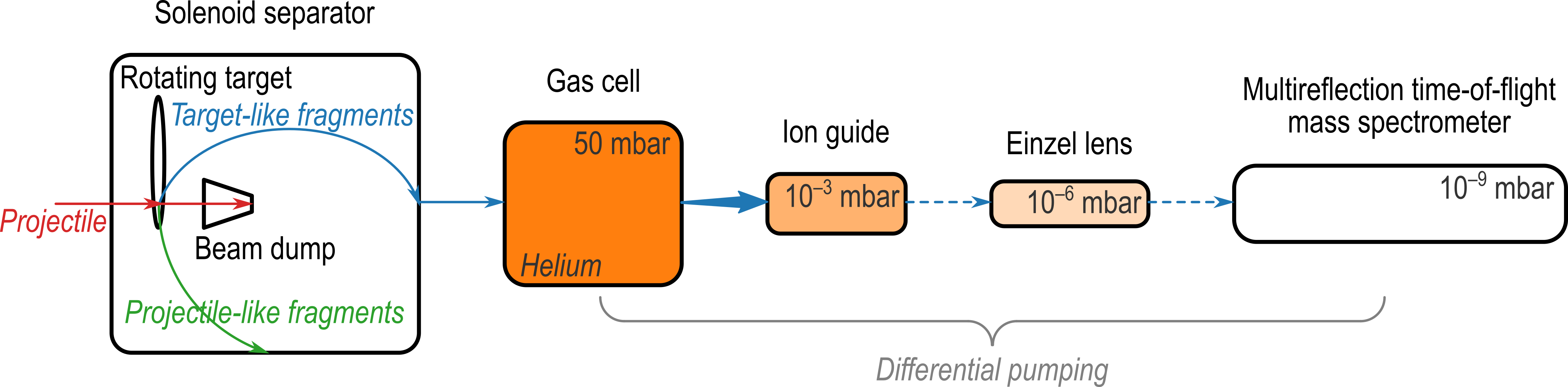}
    \caption{
    Schematic overview of the NEXT setup.
    The red arrows indicate the trajectory of a projectile.
    The green arrow shows the trajectories of projectile-like fragments, while the blue arrows show the trajectories of target-like fragments.
    The solid lines represent a continuous beam while the dashed lines represent a bunched beam.
    The helium pressure of each differential pumping stage is also labeled on the figure.
    }
    \label{fig:next_setup}
\end{figure}

The present work is preformed within the framework of the NEXT project~\cite{even_next}, which aims to measure the masses of Neutron-rich EXotic nuclei produced in multi-nucleon Transfer reactions.
The isotopes of experimental interest are located in two regions on the nuclide chart, namely the $N=126$ isotonic chain and the $Z>100$ transfermium area.
A schematic overview of the NEXT setup is shown in Fig.~\ref{fig:next_setup}.
A rotating target is irradiated by projectile nuclei, which are delivered by the superconducting AGOR cyclotron in Groningen~\cite{brandenburg_irradiation_2007}.
The resulting target-like fragments that leave the target in the forward direction are preseparated and focused by a solenoid separator~\cite{dvorak_irisexploring_2011}.
They are then stopped in a gas cell filled with helium~\cite{mollaebrahimi_setup_2020}, from which they are extracted by a supersonic gas jet.
The ions are captured, cooled, and bunched by a novel ion guide before they are transferred via an einzel lens to a downstream Multi-Reflection Time-of-Flight Mass Spectrometer (MR-ToF MS) for isobaric separation and precision mass measurements~\cite{wolf_-line_2012}.
The ion guide needs to be able to capture a divergent and continuous beam carried by the gas jet, and convert it to focused and compact ion bunches, which can then be accepted by the MR-ToF MS.
Furthermore, the ion guide and the einzel lens form differential pumping stages, which reduce the buffer gas pressure stepwise from $50$~mbar in the gas cell to $10^{-9}$~mbar in the MR-ToF MS.
The design of the ion guide, which consists of stacked ring electrodes and utilizes a digital RF potential, is put forward in the following. 

\section{Analytical Description}
The motion of an ion with mass $m$ and charge $q$ in a harmonically oscillating electric field has well been studied by Gerlich under adiabatic approximation~\cite{gerlich_inhomogeneous_1992}.
It is a superposition of a slow secular motion $\mathbf{x}_0(t)$ and a fast wiggling motion $\mathbf{x}_1(t)$, where the former is a time-averaged motion while the latter is a harmonic motion due to the local dipolar field.
Sometimes, $\mathbf{x}_0$ and $\mathbf{x}_1$ are also referred to as a macromotion and a micromotion, respectively.
This treatment can likewise be generalized for any electric field $\mathbf{E}(\mathbf{x}_0)$ with an oscillating time dependency $w(t)$, given that $w(t)$ is periodic with a period $\tau$ and its time average over one period vanishes: $\langle w \rangle = 0$.

On the one hand, the equation of the wiggling motion reads
\begin{equation}
    m\ddot{\mathbf{x}}_1 = q\mathbf{E}(\mathbf{x}_0)w(t).
    \label{eq:wiggling_motion_equation}
\end{equation}
Its solution immediately results from direct integration:
\begin{equation}
    \mathbf{x}_1 = \frac{q\mathbf{E}(\mathbf{x}_0)}{m}u(t),\quad\mathrm{with}\quad\ddot{u}(t) = \dot{v}(t) = w(t).
    \label{eq:wiggling_motion_solution}
\end{equation}
Two free integral constants in $u(t)$ can be determined by reasonably assuming that the wiggling motion returns to its starting point after one period: $u(t+\tau)=u(t)$ for any $t$, and it extends evenly in both directions around the trajectory of the secular motion: $\hat{u} := \max[u(t)] = -\min[u(t)]$.
The latter suggests that the amplitude $\mathbf{d}$ of the wiggling motion is $q\hat{u}\mathbf{E}/m$.
Under adiabatic approximation, the field variation seen by the wiggling motion is much smaller than the local field strength.
A characteristic parameter, called adiabaticity $\eta$, is therefore defined:
\begin{equation}
    \eta = \frac{|(2\mathbf{d}\cdot\nabla)\mathbf{E}|}{E} = \frac{2q\hat{u}}{m}|\nabla E|,
    \label{eq:adiabaticity}
\end{equation}
which should be kept small under some threshold $\bar\eta$ for the validity of the adiabatic approximation.
Gerlich found that $\bar\eta = 0.3$ is suitable for most practical applications~\cite{gerlich_inhomogeneous_1992}.

On the other hand, the equation of the secular motion, after time-averaging, can be approximated to
\begin{equation}
    m\ddot{\mathbf{x}}_0 = q(\mathbf{x}_1\cdot\nabla)\mathbf{E}(\mathbf{x}_0)w(t) \simeq \nabla\left(\frac{q^2E^2}{2m}\langle uw\rangle\right).
    \label{eq:secular_motion_equation}
\end{equation}
It is evident that the secular motion is formally governed by an effective electrostatic potential, or pseudopotential $\Phi_0$:
\begin{equation}
    \Phi_0 = -\frac{qE^2}{2m}\langle uw \rangle.
    \label{eq:pseudopotential}
\end{equation}
In light of the periodicity of $u(t)$, it can be shown that the potential energy of the secular motion, being $q\Phi_0$, equals the time-averaged kinetic energy of the wiggling motion, being $q^2E^2\langle v^2\rangle/2m$, which aligns with conservation of energy.
Therefore, an ion's motion in an oscillating electric field is almost equivalent to its motion in an electrostatic field with a potential given by Eq.~(\ref{eq:pseudopotential}).

\subsection{Rectangular Waveform}
\label{sec:rectangular_waveform}
Without loss of generality, we assume that $w(t)$ is a 2-state rectangular function with a variable duty cycle---the proportion of the high state in a period---slightly different from $50\%$, as specified by a small deviation $\delta$.
For one period it can be written as
\begin{equation}
    w(t;\delta) = \begin{cases}
    1 & 0 \leq t < \frac{1+\delta}{2}\tau, \\
    -1 & \frac{1+\delta}{2}\tau \leq t < \tau,
    \end{cases}\quad\mathrm{with}\quad|\delta|\ll 1.
    \label{eq:rectangular_wave_w}
\end{equation}
Due to the imbalance between the positive and negative duration, the wiggling motion, which is originally determined by Eq.~(\ref{eq:wiggling_motion_equation}), will effectively receive a gentle kick in every cycle.
These kicks add up coherently and result in a slow drift.
Within the context of superposition, this drifting component should belong to the secular motion rather than the wiggling motion.
As a result, the rectified wiggling motion is determined by a shifted rectangular function $\tilde{w}$:
\begin{equation}
    \tilde{w}(t;\delta) = \begin{cases}
    1-\delta & 0 \leq t < \frac{1+\delta}{2}\tau, \\
    -(1+\delta) & \frac{1+\delta}{2}\tau \leq t < \tau,
    \end{cases}
    \label{eq:rectified_rectangular_wave_w}
\end{equation} 
while the rectified secular motion is additionally influenced by an electrostatic field $\tilde{\mathbf{E}}=\delta\mathbf{E}$.

After integration of $\tilde{w}$ with respect to $t$ and proper selection of integral constants, $\tilde{u}$ is obtained, to a first-order approximation, to be
\begin{equation}
    \tilde{u}(t;\delta) = \begin{cases}
    \frac{1-\delta}{2}t^2 - \frac{\tau}{4}t + \frac{\delta\tau^2}{32} & 0 \leq t < \frac{1+\delta}{2}\tau, \\
    -\frac{1+\delta}{2}t^2 + \frac{(3+4\delta)\tau}{4}t - \frac{(8+15\delta)\tau^2}{32} & \frac{1+\delta}{2}\tau \leq t < \tau.
    \end{cases}
    \label{eq:rectified_rectangular_wave_u}
\end{equation}
Plugging Eqs.~(\ref{eq:rectified_rectangular_wave_w}) and (\ref{eq:rectified_rectangular_wave_u}) into Eqs.~(\ref{eq:adiabaticity}) and (\ref{eq:pseudopotential}) and neglecting higher order terms give rise to the first-order approximations of the adiabaticity and pseudopotential for the digital RF potential of the rectangular waveform:
\begin{align}
    \eta &= \frac{q|\nabla E|}{16mf^2}, \label{eq:rectangular_wave_adiabaticity} \\
    \Phi_0 &= \frac{qE^2}{96mf^2}, \label{eq:rectangular_wave_pseudopotential}
\end{align}
where $f = 1/\tau$ is the RF frequency.
The $\delta$-independent formulas in Eqs.~(\ref{eq:rectangular_wave_adiabaticity}) and (\ref{eq:rectangular_wave_pseudopotential}) reveal that for a small deviation of the duty cycle from $50\%$, the rectified wiggling motion is almost unaffected while the rectified secular motion is subject to a proportionally small electrostatic field $\delta\mathbf{E}$ apart from the pseudopotential $\Phi_0$.
This observation entails that by adjusting the duty cycle of the digital RF, the DC level of the RF potential is accordingly biased.

\subsection{Stacked Rings}
\label{sec:stacked_rings}
Owing to the rotational symmetry of stacked rings, the interior electrostatic potential $\Phi$ is preferably expressed in cylindrical coordinates $(r,\phi,z)$, where the $z$-direction aligns with the central axis of the concentric rings.
In general, the expression involves multiple terms, also known as modes, which are determined by a boundary condition~\cite{heerens_solution_1976}.
Nevertheless, the simplest fundamental mode can be written as
\begin{equation}
    \Phi(r,z) = A I_0(kr) \sin(kz),
    \label{eq:electrostatic_potential}
\end{equation}
where $A$ and $k$ are scaling factors, and $I_0$ is the modified Bessel function of the first kind $I_\nu$ with $\nu=0$.

\begin{figure}[htbp]
    \centering
    \includegraphics[width=.6\textwidth]{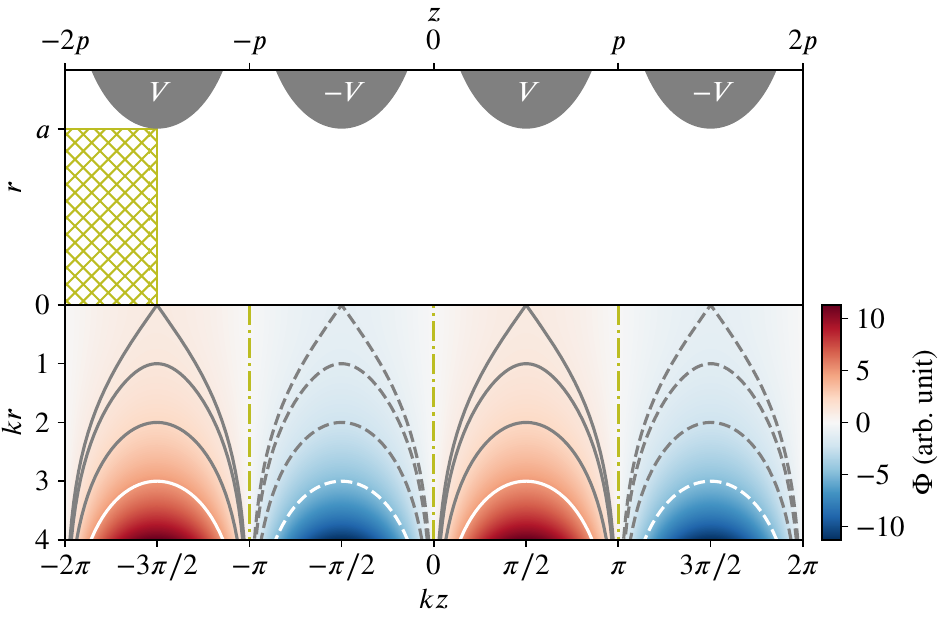}
    \caption{
    (Lower) electrostatic potential map of the fundamental mode in cylindrical coordinates with a rotational symmetry.
    The solid and dashed lines denote the positive and negative equipotential lines, respectively.
    The dash-dotted lines indicate a null potential.
    (Upper) cross-sectional view of ring electrodes that build up this potential with given voltages $\pm V$.
    The hatched box encloses a potential-unique region, from which the potential elsewhere can be derived by charge conjugation and/or parity inversion.
    }
    \label{fig:potenial_map_electrode_shape}
\end{figure}

A color-coded map of $\Phi(r,z)$ is plotted on the lower panel of Fig.~\ref{fig:potenial_map_electrode_shape}, together with equipotential lines to better illustrate the trend.
In particular, null-potential lines lie in the radial direction, as shown by the dash-dotted lines.
Such a potential can be induced by aligning the surfaces of charged rings to certain equipotential lines, such as the bright ones on the lower panel, or equivalently, the corresponding ones on the upper panel.
Note that the voltages of adjacent rings have the same magnitude $V$ but opposite polarities.
As a result, the potential in the inner volume can be represented by a small fraction in a subvolume---the potential-unique region as shown by the hatched box on the upper panel.
It is trivial to derive the potential elsewhere by charge conjugation and/or parity inversion.
Also labeled on the upper panel are the radius $a$ of the aperture and the pitch $p$ of the rings, where the latter is the distance between two adjacent null-potential lines.
Hence, the scaling factors $A$ and $k$ in Eq.~(\ref{eq:electrostatic_potential}) can be written as $V/I_0(ka)$ and $\pi/p$, respectively.

\begin{figure}[htbp]
    \centering
    \includegraphics[width=.6\textwidth]{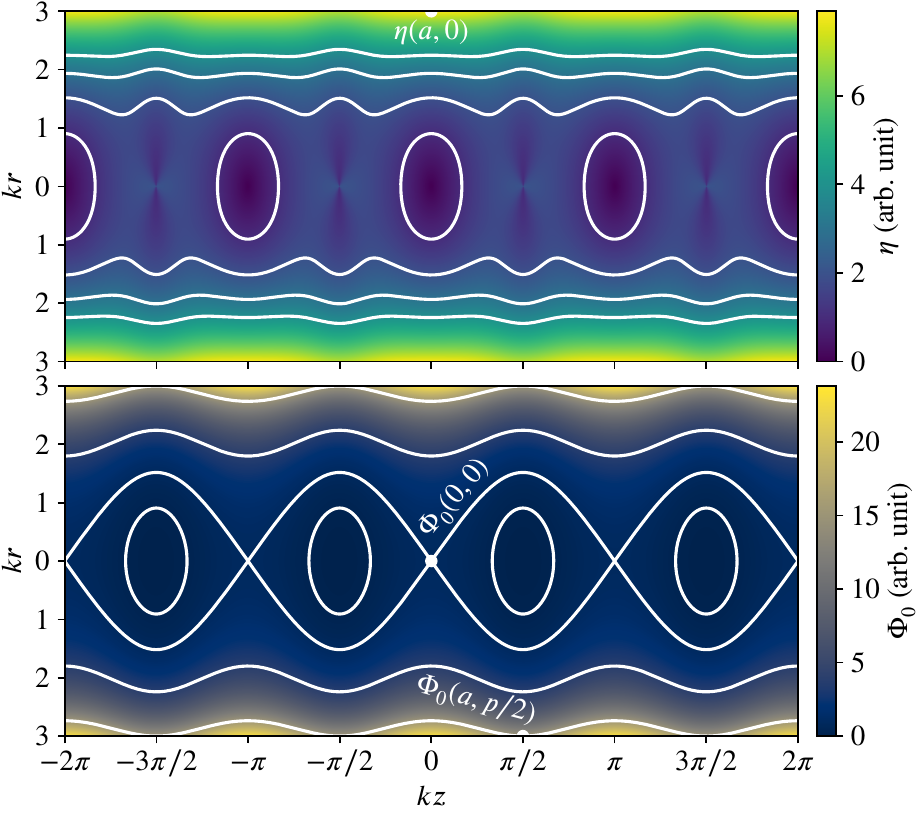}
    \caption{
    Adiabaticity map (upper) and pseudopotential map (lower) inside the stacked rings.
    The contours of each map are shown by the solid lines.
    Both maps adopt dimensionless coordinates, while at some locations the values are labeled in the corresponding physical coordinates.
    }
    \label{fig:adiabaticity_pseudopotential_map}
\end{figure}

The electric field $\mathbf{E}$ inside stacked rings can be obtained by taking the gradient of the potential $\Phi$:
\begin{align}
    \begin{split}
        E_r(r,z) &= -\frac{Vk}{I_0(ka)}I_1(kr)\sin(kz), \\
        E_z(r,z) &= -\frac{Vk}{I_0(ka)}I_0(kr)\cos(kz).
    \end{split}
    \label{eq:electrostatic_field}
\end{align}
The adiabaticity $\eta$ and the pseudopotential $\Phi_0$ can be obtained by plugging Eq.~(\ref{eq:electrostatic_field}) into Eqs.~(\ref{eq:rectangular_wave_adiabaticity}) and (\ref{eq:rectangular_wave_pseudopotential}), respectively.
Color-coded maps of $\eta(r,z)$ and $\Phi_0(r,z)$ are plotted in Fig.~\ref{fig:adiabaticity_pseudopotential_map}, together with their respective contours.

Recalling the constraint on the adiabaticity, it follows that $\max(\eta)=\eta(a,0)$ should satisfy the following inequality:
\begin{equation}
    \bar\eta \geq \eta(a, 0) = \frac{qVk^2I_1(ka)}{16mf^2I_0(ka)}.
    \label{eq:adiabaticity_constraint}
\end{equation}
To confine an ion in the radial direction, the pseudopotential barrier on the boundary, which is represented by $\Phi_0(a,p/2)$, must be higher than its initial kinetic energy $E_i$:
\begin{equation}
    E_i \leq q\Phi_0(a, p/2) = \frac{q^2V^2k^2I_1^2(ka)}{96mf^2I_0^2(ka)}.
    \label{eq:pseudopotential_barrier_constraint}
\end{equation}
As the ion loses its energy via ion-neutral collisions, it will move towards the central axis and may even be confined in one of the pseudopotential buckets, which are enclosed by the contours corresponding to the value of $\Phi_0(0,0)$ on the lower panel of Fig.~\ref{fig:adiabaticity_pseudopotential_map}.
This scenario forms a basis for ion bunching, which entails that the final kinetic energy $E_f$ is constrained by
\begin{equation}
    E_f \leq q\Phi_0(0,0) = \frac{q^2V^2k^2}{96mf^2I_0^2(ka)}.
    \label{eq:pseudopotential_bucket_constraint}
\end{equation}

Eqs.~(\ref{eq:adiabaticity_constraint}) to (\ref{eq:pseudopotential_bucket_constraint}) have set ground rules for determining the parameters of a stacked-ring ion guide.
Although they are derived based on a constant radius $a$ over the entire structure, the equations can still facilitate parameter selections in case of varying apertures, which is useful in practice for the manipulation of the radial beam size.
Furthermore, the pitch $p$, or equivalently $k$, is varied accordingly such that the pseudopotential barrier, being the right hand side of Eq.~(\ref{eq:pseudopotential_barrier_constraint}), stays constant on the edges of the varying apertures.
By this means, the electric boundary matches the geometric one, thereby maximizing the confining volume and reducing ion losses.

\section{Numerical Study}
The ion guide will be coupled to an existing gas cell~\cite{mollaebrahimi_setup_2020}, which is operated with helium buffer gas at a pressure of $50$~mbar (see Fig.~\ref{fig:next_setup}).
The gas cell is equipped with a DC cage along its axis and an RF carpet on its back end to transport thermalized ions to the exit pinhole of $0.45$-mm diameter at the center of the RF carpet.
Behind the gas cell, a turbomolecular pump (Leybold~MAG~W~2200) with a pumping speed of $2100$~l/s is employed to efficiently pump helium out, resulting in a much lower pressure of $10^{-3}$~mbar.

Ion extraction from the gas cell was simulated with the COMSOL Multiphysics\textsuperscript\textregistered\ software version $5.1$ and reported in Ref.~\cite{mollaebrahimi_setup_2020}.
The trajectories of the extracted ions fill up a conical volume with a wide opening angle of $34$\textdegree, while most of them are concentrated near the conical surface.
Their kinetic energies fall in a narrow range from $2.6$~eV to $2.8$~eV.
We therefore choose $3$~eV as a nominal value for the initial kinetic energy $E_i$ to reserve some design margin.

The distribution of the final kinetic energy $E_f$ is dictated by the ambient temperature $T$, since ions will eventually reach thermal equilibrium with the buffer gas after sufficient ion-neutral collisions.
According to the equipartition theorem, this energy receives equal contributions from all of its various modes, as indicated by the Degrees of Freedom (DoF, denoted by $n$).
According to Section~\ref{sec:rectangular_waveform}, an ion in an oscillating electric field will not only have the kinetic energy of the secular motion with $3$ DoFs, but also the pseudopotential energy with another $3$ DoFs, which is actually stored as the kinetic energy of the wiggling motion.
The total energy, $E_f$, must follow a chi-squared distribution $\chi_n^2$ with $n=6$:
\begin{equation}
    E_f \sim \frac{1}{2}\left(\frac{E_f}{k_BT}\right)^2 e^{-E_f/k_BT},
    \label{eq:equipartition}
\end{equation}
where $k_B$ is the Boltzmann constant.
This is a semibounded distribution with only a lower limit.
In practice, we cut off at the $99$th percentile, which results in $E_f=0.214$~eV for $T=295$~K as a nominal value for the final kinetic energy $E_f$.

\subsection{Geometric Parameters}
The ideal shape of the ring electrodes as shown in Fig.~\ref{fig:potenial_map_electrode_shape} turns out to be unrealistic, since its radial dimension is not finitely bounded and adjacent rings would in practice touch each other at large radii.
Therefore, we have adopted a simpler ring shape, of which an illustration is given in Section~\ref{sec:layout} (see Fig.~\ref{fig:electrodes_arrangement}).
All rings are annular plates with a fixed outer radius, while the inner radii vary between $2$~mm and $7$~mm.
Moreover, the sharp inner edges are radiused at $0.5$~mm to reduce risks of electric discharge.
Apart from the pitch $p$, the thickness $l$ of the plate also needs to be determined for a given inner radius $a$.
Because of limited manufacturing precisions, these geometric parameters can only take certain rounded values in practice.
Specifically, $p$ and $l$ are rounded to one-tenth of a millimeter, while $a$ is controlled with a quarter-millimeter precision.

\begin{figure}[htbp]
    \centering
    \includegraphics[width=.6\textwidth]{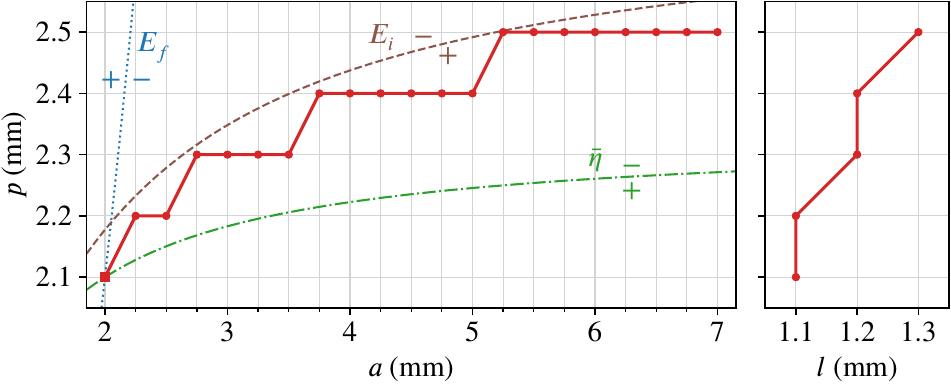}
    \caption{
    Correspondences among the radius $a$, pitch $p$ (left), and thickness $l$ (right) of realistic rings, as shown by the dots.
    The solid lines are just a guide for the eye.
    On the left panel, the dash-dotted curve is a contour of the adiabaticity with the value $\bar\eta$.
    On its either side, a ``$+$'' or ``$-$'' sign denotes the region where the adiabaticity is greater or less than $\bar\eta$.
    Likewise, the dashed and the dotted curves correspond to the initial energy $E_i$ and the final energy $E_f$, respectively.
    See text for more details.
    }
    \label{fig:geometric_parameters}
\end{figure}

The search for the optimal $p$ corresponding to a given $a$ while respecting the dimensional constraints is graphically demonstrated on the left panel of Fig.~\ref{fig:geometric_parameters}, where practically feasible $(a,p)$ pairs must locate at grid points.
The procedure begins with initial values $(a_n,p_n)=(2,2.1)$~mm, which are taken as the narrowest aperture and marked by a square dot at the bottom left corner of the plot.
An equi-adiabaticity line of $\bar\eta$, derived from Eq.~(\ref{eq:adiabaticity_constraint}), is drawn across the square dot as a dash-dotted curve on the plot, with a ``$+$'' (``$-$'') sign denoting the ascending (descending) side where the adiabaticity at any point is greater (less) than $\bar\eta$.
The inequality in Eq.~(\ref{eq:adiabaticity_constraint}) for any aperture is then translated to finding a solution on the descending side of $\bar\eta$.
Likewise, an equi-final-energy line of $E_f$, derived from Eq.~(\ref{eq:pseudopotential_bucket_constraint}), is drawn across the square dot as a dotted curve with associate ``$+$'' and ``$-$'' signs.
That ions are only bunched at the narrowest aperture requires a solution to be limited to the descending side of $E_f$.
Due to the common scaling factor on the right hand sides of Eqs.~(\ref{eq:pseudopotential_barrier_constraint}) and (\ref{eq:pseudopotential_bucket_constraint}), an equi-initial-energy line of $E_i$ is accordingly determined, which is drawn as a dashed curve on the plot with associate ``$+$'' and ``$-$'' signs.
To let the electric boundary match the geometric one while fulfilling the inequality in Eq.~(\ref{eq:pseudopotential_barrier_constraint}), the solution of $(a,p)$ pairs is eventually found to be the closest grid points to the $E_i$-line on the ascending side, which are marked by round dots on the plot.
Note that the square dot is also on the ascending side, which guarantees radial confinement for the narrowest aperture.

The optimal thickness $l$ corresponding to a given $(a,p)$ pair determined above is found by minimizing the root-mean-square deviation of an actual electrostatic potential $\Phi$ to the ideal one over the potential-unique region (see the hatched box in Fig.~\ref{fig:potenial_map_electrode_shape}), where the former is numerically solved by the SIMION\textsuperscript\textregistered\ software version $8.1$ and the latter is calculated from Eq.~(\ref{eq:electrostatic_potential}).
Interestingly, the optimal $l$ does not explicitly depend on $a$.
Its dependency on $p$ is illustrated on the right panel of Fig.~\ref{fig:geometric_parameters}.
Note that with the optimal $l$, $\Phi$ is merely distorted by a few parts per thousand, which supports the pragmatic simplification of the ideal ring shape.
As an example, the root-mean-square relative deviation of $\Phi$ is $3\times10^{-3}$ for $a=7$~mm, $p=2.5$~mm, and $l=1.3$~mm.
Furthermore, the null-potential lines (see the dash-dotted lines in Fig.~\ref{fig:potenial_map_electrode_shape}) remain intact due to the reflection symmetry of $\Phi$ in the axial direction.

\subsection{Electric Parameters}
It can be concluded from the left panel of Fig.~\ref{fig:geometric_parameters} that the adiabatic approximation is valid for all the determined $(a,q)$ pairs if Eq.~(\ref{eq:adiabaticity_constraint}) holds for $(a_n,q_n)$.
Moreover, to avoid imposing unnecessary obstacles to ion transport in the axial direction, only $(a_n,q_n)$ should fulfill Eq.~(\ref{eq:pseudopotential_bucket_constraint}).
Therefore, two electric parameters, namely the frequency $f$ and the voltage $V$ of a digital RF, can be constrained by substituting $a_n$ and $k_n=\pi/p_n$ in Eqs.~(\ref{eq:adiabaticity_constraint}) and (\ref{eq:pseudopotential_bucket_constraint}).
Canceling out $V$ in both equations results in the constraint on $f$:
\begin{equation}
    f \geq \frac{k_nI_1(k_na_n)}{4\bar\eta}\sqrt{\frac{6E_f}{m}}.
    \label{eq:digital_rf_frequency}
\end{equation}
Plugging this constraint in Eq.~(\ref{eq:pseudopotential_bucket_constraint}) gives rise to
\begin{equation}
    V \geq \frac{6I_0(k_na_n)I_1(k_na_n)E_f}{\bar\eta q}.
    \label{eq:digital_rf_voltage}
\end{equation}
As an example, to cool and bunch $^{203}\mathrm{Ir}^+$ ions in the NEXT experiments, the minimum frequency and voltage of the digital RF are $3.82$~MHz and $82$~V, respectively.

Eqs.~(\ref{eq:digital_rf_frequency}) and (\ref{eq:digital_rf_voltage}) reveal scaling laws of electric parameters with respect to ion properties.
For ions extracted from a gas cell, which have a common charge state of $1+$, or $2+$ for superheavy elements, the voltage is independent of the particular ion species.
The frequency, however, is inversely proportional to the square root of the ion mass.
Therefore, a stacked-ring ion guide designed for given initial and final kinetic energies can serve as a high-pass mass filter with fixed frequency and voltage.

\section{Conceptual Design}
The ion guide should have the widest aperture at the entrance to efficiently capture the aforementioned incoming beam with a divergent emitting angle.
At the exit, the aperture should be the narrowest to produce well focused ion bunches.
The entire structure should serve the purpose of ion cooling and bunching while maintaining a decent transmission efficiency.
Ion trajectory simulations are performed with SIMION\textsuperscript\textregistered\ to optimize the design with respect to these objectives.
For the incoming beam, $300$ randomly sampled $^{203}\mathrm{Ir}^+$ ions are used as test particles in the simulations.
The ion guide is operated with helium buffer gas at a pressure of $10^{-3}$~mbar.
The corresponding ion-neutral collisions are implemented according to an elastic hard-sphere model by the SIMION\textsuperscript\textregistered\ library \verb|collision_hs1|.
Table~\ref{tab:conditions} summarizes the circumstances under which the design has been optimized.

\begin{table}[htbp]
    \centering
    \caption{
    List of conditions for the design optimization of the ion guide.
    }
    \begin{tabular}{l c r}
        \hline
        Description & Symbol & Value \\
        \hline
        Buffer gas & & He \\
        Pressure & & $10^{-3}$~mbar \\
        Temperature & $T$ & $295$~K \\
        Upper limit of adiabaticity & $\bar\eta$ & $0.3$ \\
        Nominal final energy & $E_f$ & $0.214$~eV \\
        Divergence of incoming beam & & $34$\textdegree \\
        Nominal initial energy & $E_i$ & $3$~eV \\
        Test ions & & $^{203}\mathrm{Ir}^+$ \\
        Number of ions & & $300$ \\
        \hline
    \end{tabular}
    \label{tab:conditions}
\end{table}

\subsection{Layout}
\label{sec:layout}
\begin{figure}[htbp]
    \centering
    \includegraphics[width=\textwidth]{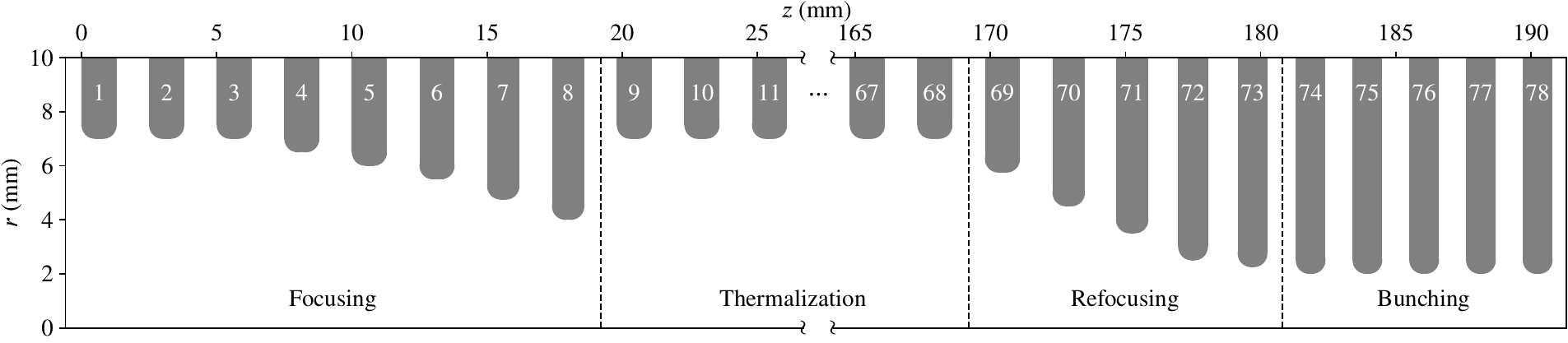}
    \caption{
    Cross-sectional view of the ion guide in an actual scale as indicated by the coordinates.
    The total $78$ constituent rings are numbered sequentially from entrance to exit.
    The inner edges of each ring are radiused at $0.5$~mm.
    The ion guide consists of four functional sections, which are labeled in the figure.
    }
    \label{fig:electrodes_arrangement}
\end{figure}

\begin{figure}[htbp]
    \centering
    \includegraphics[width=.75\textwidth]{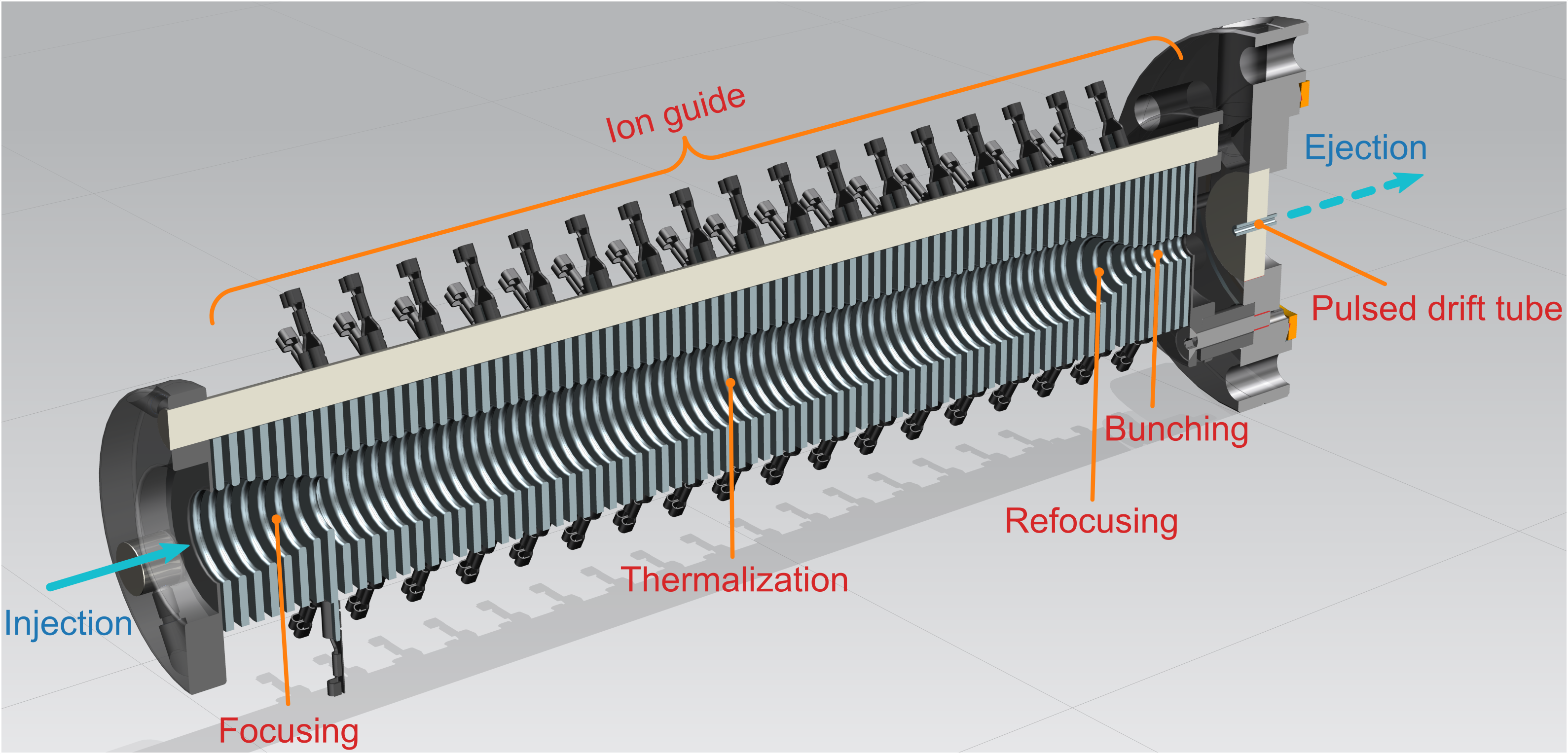}
    \caption{
    Cutaway drawing of the three-dimensional model of the ion guide. 
    It comprises focusing, thermalization, refocusing, and bunching sections. 
    A divergent and continuous beam coming from left is converted to focused and compact ion bunches going to right.
    A pulsed drift tube behind the ion guide accelerates the ejected ion bunches.
    }
    \label{fig:3d_model}
\end{figure}

The ion guide consists of $78$ concentrically stacked rings of varying apertures.
The layout is schematically shown in Fig.~\ref{fig:electrodes_arrangement}, where every constituent ring is numbered sequentially from entrance to exit, and quantitatively listed in Table~\ref{tab:geometric_parameters}.
Additionally, a three-dimensional model of the ion guide is shown in Fig.~\ref{fig:3d_model}.
To restrain the field distortion due to the aperture change, all the rings are laid out in such a way that the null-potential lines of a ring overlap with those of its neighbors.
Recalling the definition of the pitch in Section~\ref{sec:stacked_rings}, the total length of the ion guide, which amounts to $190.8$~mm, is thus the sum of all the pitches except for the rings at both ends.
According to the right panel of Fig.~\ref{fig:geometric_parameters}, the thickness of a ring is always less than its pitch.
Therefore, isolating gaps can naturally arise in between.

\begin{table}[htbp]
    \centering
    \caption{
    Geometric parameters of the ion guide.
    }
    \begin{tabular}{l c r}
        \hline
        Description & Symbol & Value \\
        \hline
        Number of rings & & $78$ \\
        Total length & & $190.8$~mm \\
        Radii of apertures & $a$ & $2$~mm to $7$~mm \\
        Pitches of rings & $p$ & $2.1$~mm to $2.5$~mm \\
        Thicknesses of rings & $l$ & $1.1$~mm to $1.3$~mm \\
        Radius of inner edges & & $0.5$~mm \\
        Outer radius & & $10$~mm \\
        \hline
    \end{tabular}
    \label{tab:geometric_parameters}
\end{table}

The specific shape of the ion guide mainly stems from the optimization of the transmission efficiency.
With the helium buffer gas at a pressure of $10^{-3}$~mbar, the mean-free-path of test ions is around $5$~mm.
An ion's trajectory within the ion guide is predominantly determined by electric fields while ion-neutral collisions only impose a light damping effect.
Consequently, incoming ions tend to be reflected right away by contracting apertures, which are necessary for the transition from the wide entrance to the narrow exit.
As a remedy, a spacious thermalization section is included in the ion guide to sufficiently cool the ions before they are transported towards the exit.
The entire structure is functionally divided into four parts, namely for focusing, thermalization, refocusing, and bunching.
The divergent ions coming from left first confront the focusing section and are focused to the center of the last ring (\textnumero~$8$) of this section with a-few-millimeter-wide aberration.
Afterwards, the ions enter the thermalization section, which consists of $60$ identical rings with the largest inner radius of $7$~mm.
Here, the ions will reach the thermal equilibrium with the buffer gas after sufficient ion-neutral collisions.
The thermalized ions will then be collimated by the refocusing section into the last bunching section, where the smallest inner radius of $2$~mm helps store the ions in a pseudopotential bucket to form an ion bunch.

\subsection{Ion Transport}
\begin{figure}[htbp]
    \centering
    \includegraphics[width=.7\textwidth]{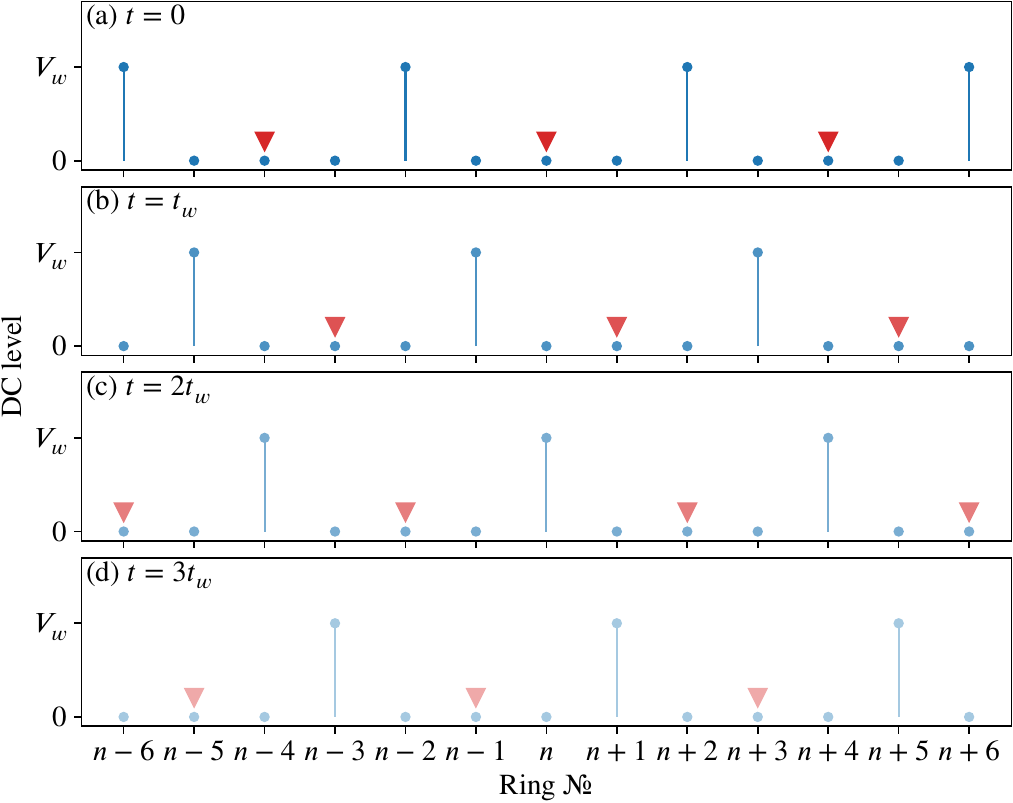}
    \caption{
    Schematic of DC-level configurations of the traveling wave for ion transport at (a) $t=0$, (b) $t=t_w$, (c) $t=2t_w$, and (d) $t=3t_w$.
    The bias voltages of the rings are indicated by the dots, while the corresponding potential wells are indicated by the triangles.
    See text for more details.
    }
    \label{fig:traveling_wave}
\end{figure}

After ions are fully thermalized in the thermalization section, the diffusion in the buffer gas can greatly delay their arrivals at the bunching section.
Thus, a traveling wave of bias voltages is employed to speed up this process.
Similar ideas have been applied in ion mobility separation~\cite{giles_applications_2004,shvartsburg_fundamentals_2008,hamid_characterization_2015,campuzano_historical_2019} and ion transport in a gas cell~\cite{colburn_ion_2004,bollen_ion_2011,brodeur_experimental_2013,brodeur_traveling_2013}.
Specifically, the wave is running from the first ring (\textnumero~$9$) of the thermalization section to the first ring (\textnumero~$74$) of the bunching section.
The schematic of all four phases of the wave is shown in Fig.~\ref{fig:traveling_wave}.
At a given moment, e.g. $t=0$, the DC level of every fourth ring is raised to a wave voltage $V_w$ while the DC levels of the others remain intact, as illustrated by the dots in Fig.~\ref{fig:traveling_wave}(a).
This configuration creates a potential well in the middle of each segment, as marked by the triangles in the figure.
The ions will be attracted to, and eventually stored in, these potential wells.
At a later moment $t=t_w$, the configuration advances towards the exit by one ring, as illustrated in Fig.~\ref{fig:traveling_wave}(b).
The thermalized ions will be pushed forward by the potential switch, but will still be confined in the same segments bouncing back and forth until they are recooled by the buffer gas towards the new potential wells, as revealed by simulations.
Consequently, the ions ``ride'' the traveling wave to the bunching section.
The optimal values of $V_w$ and $t_w$ are found by simulations to be $2.5$~V and $750$~$\mu$s, respectively.
They are compiled in Table~\ref{tab:electric_parameters}, which lists the electric parameters of the ion guide.

\begin{table}[htbp]
    \centering
    \caption{
    Electric parameters of the ion guide for radially confining and axially transporting ions.
    }
    \begin{tabular}{l c r}
        \hline
        Description & Symbol & Value \\
        \hline
        Frequency of digital RF & $f$ & $3.82$~MHz \\
        Voltage of digital RF & $V$ & $82$~V \\
        Quarter period of traveling wave & $t_w$ & $750$~$\mu$s \\
        Wave voltage & $V_w$ & $2.5$~V \\
        Corresponding deviation of duty cycle & $\delta_w$ & $3\%$ \\
        Reflection voltage & $V_r$ & $1.3$~V \\
        Corresponding deviation of duty cycle & $\delta_r$ & $1.6\%$ \\
        Gating voltage & $V_g$ & $2.5$~V \\
        Corresponding deviation of duty cycle & $\delta_g$ & $3\%$ \\
        \hline
    \end{tabular}
    \label{tab:electric_parameters}
\end{table}

\subsection{Ion Bunching}
Due to the force exerted by the traveling wave, ions entering the bunching section will initially be more energetic than any pseudopotential bucket in the section.
To avoid their immediate escapes from the exit, the DC level of the last ring (\textnumero~$78$) of the bunching section is raised to a gating voltage $V_g$.
The value of $V_g$ is set to be that of $V_w$ such that when the traveling wave propagates furthest to the ring \textnumero~$74$, a potential well is formed in the middle, i.e.\ at the ring \textnumero~$76$. 
The ions will be attracted to, and eventually stored in, this potential well.
Even after the DC level of the ring \textnumero~$74$ is restored to zero, the gating voltage on the ring \textnumero~$78$ can impose a negligible perturbation on the ions.
They will still be stored in the pseudopotential bucket at the ring \textnumero~$76$.
Consequently, the incoming ions will continually be transported to and accumulated at this location, where they form an ion bunch.

The gating voltage at the exit can also help prevent direct passages of paraxially incoming ions through the ion guide.
Moreover, based on simulations, an incoming ion can sometimes bounce back in the refocusing section.
To reflect such backwards flying ions that are still to be thermalized, the DC level of the last ring (\textnumero~$8$) of the focusing section is raised to a reflection voltage $V_r$.
The optimal value of $V_r$ without hampering ion injections into the thermalization section is found by simulations to be $1.3$~V.
Although it is less than the value of $V_w$ (see Table~\ref{tab:electric_parameters}), ion reflections caused by $V_r$ is enhanced by a smaller aperture of the ring \textnumero~$8$ than that of the rings in the thermalization section.

\subsection{Ion Ejection}
When an ion bunch is ready for ejection, the last four rings (\textnumero~$75$ to \textnumero~$78$) of the ion guide will be disconnected from the digital RF and independently charged to ejection voltages.
The voltage change can be completed typically in less than $40$~ns with a fast switch such that the ion bunch will virtually experience no loss of confinement before it is accelerated by the ejection gradient.
Special attention ought to be paid to the timing of the ejection.
It should occur right before the DC level of the ring \textnumero~$74$ is restored to zero.
As a result, the ion bunch is already in the thermal equilibrium, and the least perturbation is imposed on other ions that are being transported.
The ensemble of the ejection rings acts as an ion-optical lens system, which can flexibly focus the ejected ion bunch on the axis towards a given point with corresponding ejection voltages.
This is particularly useful when the ion bunch needs to pass through a tiny orifice to the downstream higher vacuum stage.
Table~\ref{tab:ejection_parameters} gives an example of optimal ejection voltages to focus the ion bunch $12$ mm behind the ion guide.

\begin{table}[htbp]
    \centering
    \caption{
    Ejection voltages applied to the last four rings of the ion guide.
    }
    \begin{tabular}{l r}
        \hline
        Description &  Value \\
        \hline
        Focal length & $12$~mm \\
        Ejection voltage for ring \textnumero~$75$ & $140$~V \\
        Ejection voltage for ring \textnumero~$76$ & $100$~V \\
        Ejection voltage for ring \textnumero~$77$ & $95$~V \\
        Ejection voltage for ring \textnumero~$78$ & $85$~V \\
        \hline
    \end{tabular}
    \label{tab:ejection_parameters}
\end{table}

\subsection{Realization}
The realization of the entire process described above critically depends on the digital RF that is applied to the ion guide.
It is typically generated by two Power Supply Units (PSUs) with the same magnitude $V$ but opposite polarities and two fast switches.
The latter switch between the positive and negative PSU at the RF frequency $f$, and feed two alternating voltages respectively to even and odd rings.

\begin{figure}[htbp]
    \centering
    \includegraphics[width=.6\textwidth]{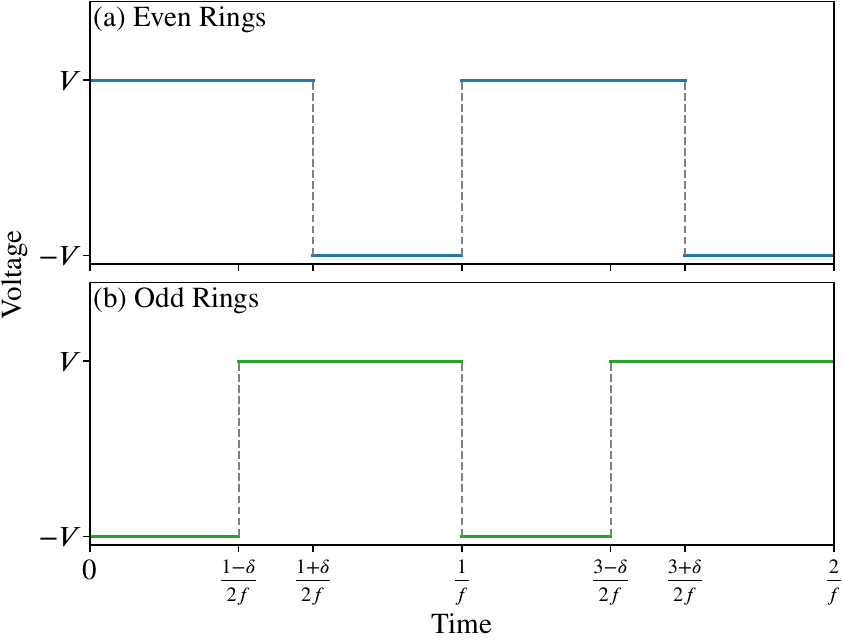}
    \caption{
    Time spectra of the confining digital RF for (a) even and (b) odd rings.
    Owing to the positive deviation $\delta$ of the duty cycle from $50\%$, both spectra effectively induce the same bias voltage of $\delta V$.
    }
    \label{fig:time_spectrum}
\end{figure}

Such a scheme has readily been applied to digitally driven RFQs and ion funnels.
However, for the present ion guide, the DC levels of certain rings need to be individually raised to bias voltages, as listed in Table~\ref{tab:electric_parameters}.
According to Section~\ref{sec:rectangular_waveform}, such a voltage, e.g.\ $\delta V$, can effectively be induced by deviating the duty cycle of the digital RF from $50\%$ by an amount of $\delta$.
The resultant time spectra for even and odd rings in case of a positive deviation $\delta$ are shown in Fig.~\ref{fig:time_spectrum}.
The induced bias voltage can easily be adjusted by reprogramming the timing pattern for the fast switch without any modifications to the hardware.

\begin{figure}[htbp]
    \centering
    \includegraphics[width=.6\textwidth]{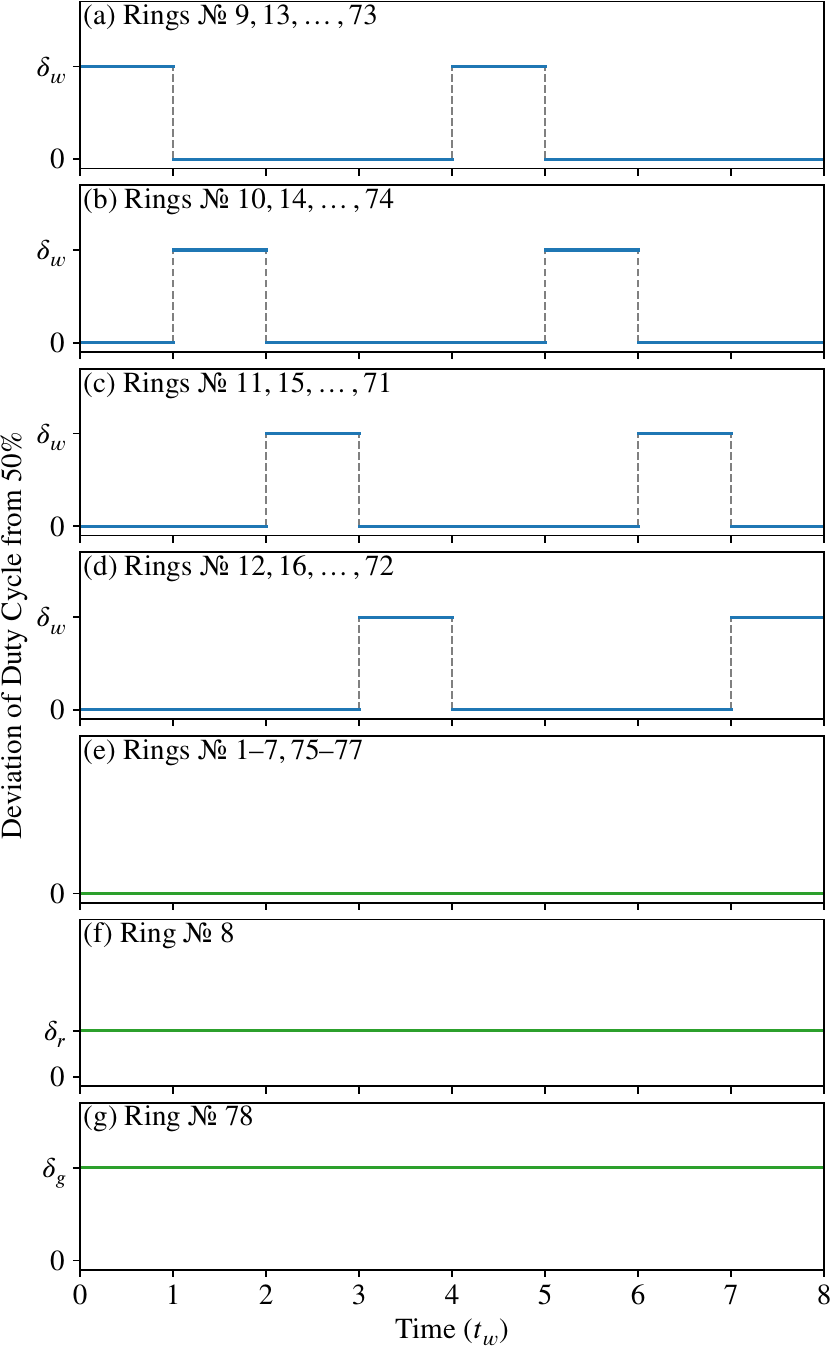}
    \caption{
    Timing pattern of the deviation of the duty cycle from $50\%$ during ion cooling and bunching for every constituent ring of the ion guide.
    }
    \label{fig:timing_diagram}
\end{figure}

We note that Fig.~\ref{fig:time_spectrum} dose not necessarily imply the same spectrum being applied to all the even (or odd) rings at the same time.
During the operation of the ion guide, the duty cycle of the digital RF changes according to the phase sequence in Fig.~\ref{fig:traveling_wave} for the participating rings in the traveling wave, while it remains constant for the others.
Fig.~\ref{fig:timing_diagram} shows the timing pattern of the deviation $\delta$ of the duty cycle for every constituent ring of the ion guide.
The value of $\delta$ is calculated as the ratio of the bias voltage to the PSU voltage, and listed in Table~\ref{tab:electric_parameters} for all possible scenarios.
Based on the timing diagram in Fig.~\ref{fig:timing_diagram}, rings sharing the same pattern and polarity can directly be wired, thus notably simplifying the circuitry complexity.

\section{Expected Performance}
The present design of the ion guide has systematically been optimized via simulations with SIMION\textsuperscript\textregistered\ in various aspects, such as ion injection, transport, bunching, and ejection.
To showcase the expected performance of the ion guide, two characteristic properties, namely transmission efficiency and beam emittances, are listed in Table~\ref{tab:performance} and will be detailed in the following.

\begin{table}[htbp]
    \centering
    \caption{
    Expected performance of the ion guide according to simulations with SIMION\textsuperscript\textregistered.
    }
    \begin{tabular}{l c r}
        \hline
        Description & Symbol & Value \\
        \hline
        Transmission efficiency & & $80(3)\%$ \\
        Transverse emittance in $x$-direction & $\epsilon_{2\sigma,x}$ & $20.8(5)$~$\pi$~mm~mrad \\
        Transverse emittance in $y$-direction & $\epsilon_{2\sigma,y}$ & $21.0(5)$~$\pi$~mm~mrad \\
        Longitudinal emittance & $\epsilon_{2\sigma,z}$ & $0.9(1)$~$\pi$~$\mu$s~eV \\
        \hline
    \end{tabular}
    \label{tab:performance}
\end{table}

\subsection{Transmission Efficiency}
After sequentially injecting the aforementioned $300$ test particles into the ion guide, the full trajectory of each ion is tracked in the simulation until a terminating condition is met.
The statistics of ion terminations are listed in Table~\ref{tab:ion_termination}.
Only a fraction of ions are lost, mainly due to slipping off the traveling wave.
Most ions are successfully stored in the pseudopotential bucket at the ring \textnumero~$76$ for ion bunching.
The simulated transmission efficiency is $80\%$.
To estimate its uncertainty, the process has been repeated $10$ times.
The standard deviation of the $10$ resultant transmission efficiencies is calculated to be $3\%$, which is also included in Table~\ref{tab:performance}.

\begin{table}[htbp]
    \centering
    \caption{
    Statistics of ion terminations in a complete run of SIMION\textsuperscript\textregistered\ simulations.
    }
    \begin{tabular}{l r}
        \hline
        Termination & Count \\
        \hline
        Stored in the bunching bucket & 240 \\
        Slipping off the traveling wave & 47 \\
        Denied entry & 7 \\
        Direct passage & 2 \\
        Hitting the boundary & 2 \\
        Escaping from the bunching bucket & 2 \\
        \hline
    \end{tabular}
    \label{tab:ion_termination}
\end{table}

\subsection{Beam Emittances}
Owing to the stochastic nature of ion-neutral collisions, the thermalized ions possess no ``memories'' of their initial conditions.
Therefore, the ion bunch at the ring \textnumero~$76$ can equivalently evolve from standstill ions in the corresponding pseudopotential bucket.
To randomly draw a sample from the ion bunch, a standstill ion is released from the center of the pseudopotential bucket and its kinetic state is recorded after a random duration, which is sufficiently long for thermalization.
This process has been repeated $300$ times.

\begin{figure}[htbp]
    \centering
    \includegraphics[width=.6\textwidth]{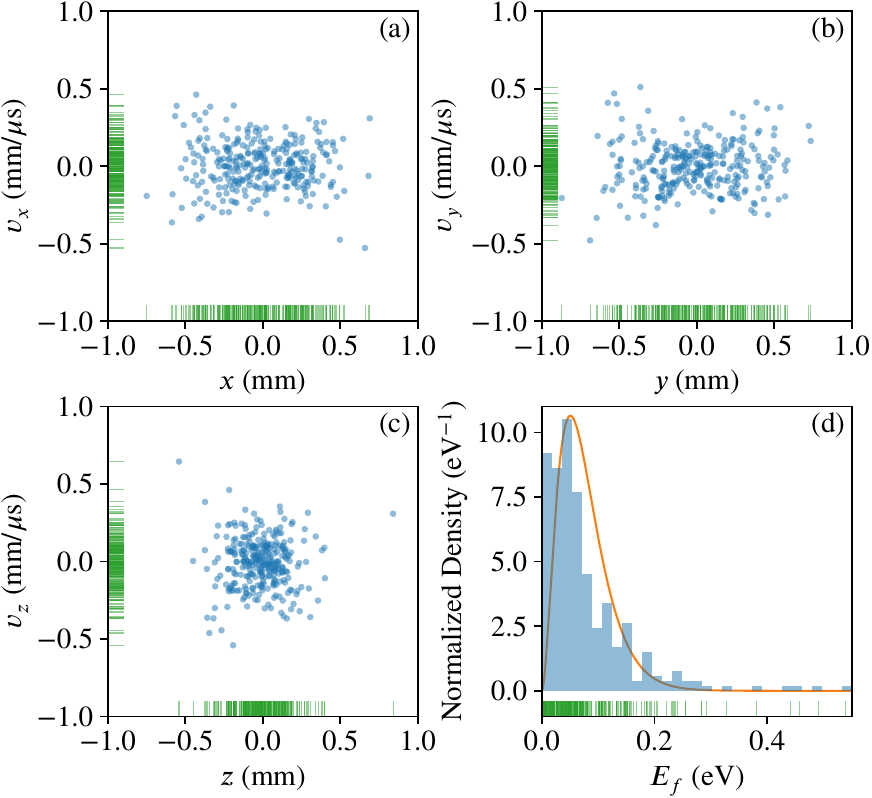}
    \caption{
    (a) $x$-related two-dimensional phase-space distribution (dots) of thermalized ions in the bunching bucket.
    Two corresponding marginal distributions (sticks) are also shown on the sides.
    (b) and (c) are similar to (a), but for $y$- and $z$-directions, respectively.
    (d) Binned (bars) and unbinned (sticks) energy distributions of the same ions.
    The theoretical prediciton is shown by the solid line.
    }
    \label{fig:standby_ions_profile}
\end{figure}

The distribution of the $300$ samples in a six-dimensional phase space is illustrated in Fig.~\ref{fig:standby_ions_profile}(a)--(c) with projections onto two-dimensional planes corresponding to three mutually orthogonal $x$-, $y$-, and $z$-directions, where the $z$-direction aligns with the axis of the ion guide.
Owing to the narrowest aperture in the bunching section, the ion bunch can loosely be confined in a ($2\times2\times2$)-mm$^3$ cubic volume of the configuration subspace, with a slightly stronger compression in the axial direction.
In the velocity subspace, the ion bunch manifests no preferences towards any direction as it is all tightly bounded within $1$-mm/$\mu$s range in each dimension.
Moreover, the energy distribution of the ion bunch shown in Fig.~\ref{fig:standby_ions_profile}(d) agrees well with the theoretical prediction given by Eq.~(\ref{eq:equipartition}), which is drawn as the solid line in the figure.

\begin{figure}[htbp]
    \centering
    \includegraphics[width=.6\textwidth]{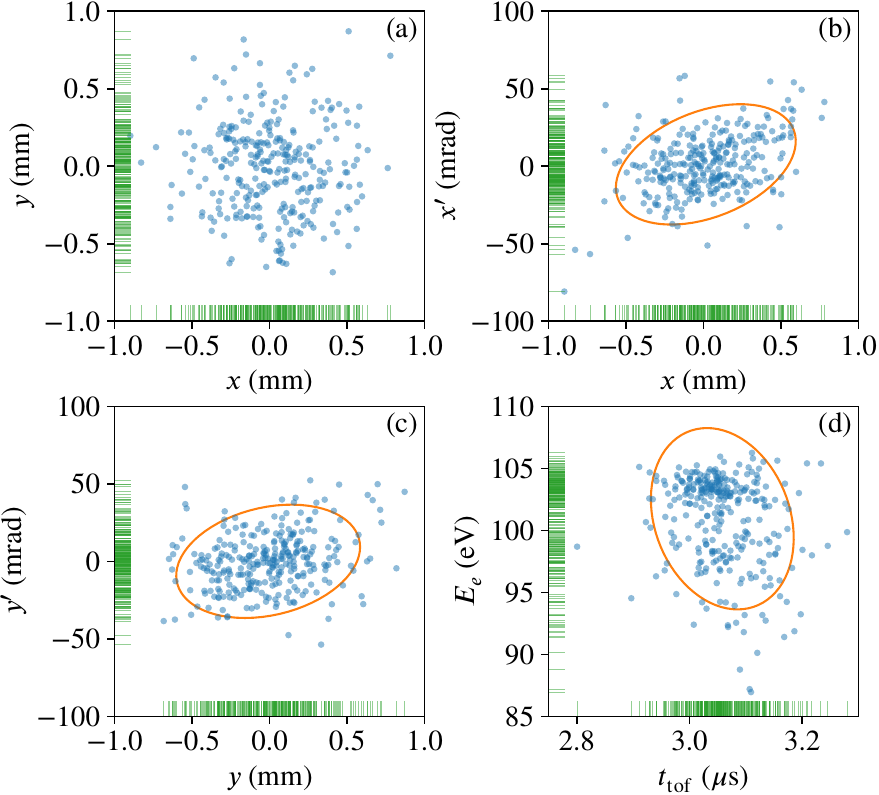}
    \caption{
    (a) Transverse beam spot (dots) of an ejected ion bunch on the focal plane and the corresponding marginal distributions (sticks) in $x$- and $y$- directions.
    (b) $x$-related two-dimensional trace-space distribution (dots) and two corresponding marginal distributions (sticks) of the ion bunch.
    The $2\sigma$ RMS $x$-emittance is illustrated by the area of the ellipse.
    (c) is similar to (b), but for $y$-direction.
    (d) Two-dimensional longitudinal phase-space distribution (dots) and two corresponding marginal distributions (sticks) of the same ion bunch.
    }
    \label{fig:ejected_ions_profile}
\end{figure}

After the ion bunch is ejected by applying the ejection voltages in Table~\ref{tab:ejection_parameters}, it is profiled on the focal plane $12$-mm behind the ion guide.
Its distribution in a four-dimensional transverse trace space---a cousin of the phase space with the velocity subspace replaced by the angle subspace with respect to the beam direction---is illustrated in Fig.~\ref{fig:ejected_ions_profile}(a)--(c) with three two-dimensional projections.
Fig.~\ref{fig:ejected_ions_profile}(a) reveals a well focused beam spot with an about $1$-mm radial size.
The root-mean-square (RMS) transverse beam emittances $\epsilon_{2\sigma,x}$ and $\epsilon_{2\sigma,y}$ at twice standard deviation ($2\sigma$) are calculated to be $20.8$~$\pi$~mm~mrad and $21.0$~$\pi$~mm~mrad, respectively.
These values are shown as the areas of the ellipses in Fig.~\ref{fig:ejected_ions_profile}(b) and (c).
Moreover, the distribution of the ion bunch in a two-dimensional longitudinal phase space is shown in Fig.~\ref{fig:ejected_ions_profile}(d), where $E_e$ is the ejection energy and $t_\mathrm{tof}$ is the time of flight of an ion to the focal plane.
The $2\sigma$ RMS longitudinal beam emittance $\epsilon_{2\sigma,z}$ is calculated to be $0.9$~$\pi$~$\mu$s~eV and is likewise illustrated in the figure.

The uncertainties of the beam emittances are obtained from $10$ repeated simulation runs in a similar manner as above.
The values corresponding to the two transverse directions both are $0.5$~$\pi$~mm~mrad, while the value for the longitudinal direction is $0.1$~$\pi$~$\mu$s~eV (see also Table~\ref{tab:performance}).
It is thus found that $\epsilon_{2\sigma,x}$ is the same as $\epsilon_{2\sigma,y}$ up to the uncertainty, which aligns with the rotational symmetry of the structure.

\section{Conclusions}
In the context of the NEXT project, a stacked-ring ion guide for cooling and bunching low-energy ions has been designed and studied by ion trajectory simulations with SIMION\textsuperscript\textregistered.
Owing to the rotational symmetry of the ion guide, the wide basin of the pseudopotential well around the central axis allows for a high ion transmission.
The apertures of the constituent rings vary over the entire structure to accept a divergent incoming beam on one end and to produce focused ion bunches on the other end.
The pitch and the thickness of each ring are varied accordingly such that the electric boundary matches the geometric one to maximize the confining volume and to reduce ion losses.
A digital RF following a rectangular waveform is generated by fast switching between a positive and a negative PSU, and is applied to the ion guide to radially confine the ions.
The RF frequency can easily be tuned to match the mass-to-charge ratio of the ions.
Moreover, the duty cycles of the digital RF on certain rings are individually deviated from $50\%$ to induce bias voltages.
A traveling wave of the bias voltages can facilitate to transport thermalized ions to the bunching section, where the ion bunch is formed.
During the operation of the ion guide, the incoming ions are continually transformed to pulsed ion bunches, which maximizes the utilizing efficiency of the beam time.
Beyond the NEXT project, this type of ion guide can manifest itself an alternative solution to the preparation of high-quality rare isotope beams in a low-energy beam line, which may optionally coupled to a high-energy separator via a gas cell, for precision measurements or post-accelerations.

\section*{Acknowledgments}
This research was funded by European Research Council Executive Agency (ERCEA), under the powers delegated by the European Commission through a starting grant number 803740---NEXT---ERC-2018-STG.

\bibliographystyle{elsarticle-num}
\bibliography{refs}

\end{document}